\documentclass[conference]{IEEEtran}
\newtheorem{definition}{Definition}[section]

\IEEEoverridecommandlockouts
\usepackage{cite}
\usepackage{amsmath,amssymb,amsfonts}
\usepackage{algorithmic}
\usepackage{graphicx}
\usepackage{textcomp}
\usepackage{xcolor}
\usepackage{mathtools}
\DeclarePairedDelimiter\ket{\lvert}{\rangle}
\def\BibTeX{{\rm B\kern-.05em{\sc i\kern-.025em b}\kern-.08em
    T\kern-.1667em\lower.7ex\hbox{E}\kern-.125emX}}
\begin{document}

\title{\title{Automated Quantum Memory Compilation with Improved Dynamic Range}
\thanks{TBD: Funding Acknowledgement}
}

\author{\IEEEauthorblockN{Aviraj Sinha, Elena R. Henderson, Jessie M. Henderson, Mitchell A. Thornton}
\IEEEauthorblockA{\textit{Darwin Deason Institute for Cyber Security, Southern Methodist University}\\
Dallas, Texas \\
\{avirajs, erhenderson, hendersonj, mitch\}@smu.edu}
}

\maketitle

\begin{abstract}
Emerging quantum algorithms that process data require that classical input data be represented as a quantum state. 
These data-processing algorithms often follow the gate model of quantum computing---which requires qubits to be initialized to a
basis state, typically $\ket{0}$---and thus often employ state generation circuits to transform the initialized basis state to a data-representation state.  
There are many ways to encode classical data in a qubit, and the oft-applied approach of basis encoding does not allow optimization to the extent that other variants do.
In this work, we thus consider automatic synthesis of addressable, quantum read-only memory (QROM) circuits, which act as data-encoding state-generation circuits. We investigate three data encoding approaches, one of which we introduce to provide improved dynamic range and precision.  We present experimental results that compare these encoding methods for QROM synthesis to better understand the implications of and applications for each.
\end{abstract}


\section{Introduction}
Quantum computing promises substantial performance gains for certain classes of algorithms, including searching databases, factoring prime numbers, and solving linear systems\cite{grover1997quantum,shor1995scheme,harrow2009quantum}.
In some cases, such as the latter two, these gains are provably exponential, and, in all cases, they are due to properties of quantum mechanics that allow quantum algorithms to behave in ways that their classical counterparts cannot.
One such property is that of superposition, which allows quantum algorithms to operate on significantly larger datasets than classical algorithms~\cite{nielsen2002quantum}.

The predominant implementation of generalized quantum computers applies the gate model of computation, as described in~\cite{feynman2018simulating} and~\cite{deutsch1985quantum}.
While such machines are theoretically capable of executing algorithms such as those described above, today's quantum computers remain noisy and fairly small in terms of the number of available qubits, given the engineering challenges inherent in building such devices.
Specifically, contemporary hardware is part of the noisy intermediate-scale quantum computing (NISQ) era, and is extremely sensitive to decoherence, necessitating that quantum algorithms employ the smallest number of qubits and quantum operations possible~\cite{preskill2018quantum}.

Usually, the first step in the specification of gate-model quantum computer (QC) algorithms is to initialize qubits to a ``ground state'' denoted as the $\ket{0}$ computational basis state~\cite{deutsch1985quantum, divincenzo2000physical}.  Many QC algorithms have a structure wherein all qubits are initialized to $\ket{0}$, evolved into a new quantum state via application of quantum gates, and measured to learn at least some of the resulting evolved state.
This computational paradigm results in significant differences between classical and quantum algorithms for processing data.
In contrast to data processing applications using a classical computer, input data must be supplied to a QC program by evolving its qubits (initialized to a basis state) into a quantum state that represents the classical data to be processed.  This sequence of processing steps is in the form of a general ``state generation'' circuit.

Many quantum algorithms, including those in the emerging field of quantum machine learning (QML), require the use of data for training, operational input, or other processing applications.
Consequently, previous work has proposed a variety of different quantum memory models, such as quantum random-access memory (QRAM) \cite{Giovannetti2008} and quantum read-only memory (QROM) \cite{babbush2018encoding}.
QROM circuits evolve a set of initialized qubits into a quantum representation of addressable, classical data, where that evolution must be unique for each unique set of classical input data.
Prior work describing QROM has addressed the important goal of efficiently performing quantum data encoding; however, given the emergence of new quantum algorithms, there remains a lack of consensus regarding both the categorization and implementation of encoding methods, as well as the standardization of QROM format for storing such encoded data sets.
This paper seeks to address that gap by presenting definitions of three different data encoding approaches, two of which have been previously discussed (basis encoding and angle encoding), and one of which is a newly-proposed method that we refer to as ``improved angle encoding.''
Additionally, we present a tool that compiles QROM circuits using these three encoding methods and corresponding encoding-specific optimizations, resulting in improved circuit design with respect to overall quantum cost and other metrics.
Specifically, our contributions are a standardized approach for synthesizing QROM circuits, including standardization of definitions; introduction of a newly proposed encoding method and optimization techniques for pre-existing encoding methods; and development of a software tool for automatically generating circuits specified as QROMs with one of three different encoding types.

The paper proceeds as follows. First, we present background information including our definition of a QROM and definitions for three quantum data encoding approaches.  Next, we clarify how our work augments and complements prior research.
We then introduce our quantum compiler tool, ``MustangQ,'' and describe how we implement three encoding methods in the automatically produced QROM structure: basis encoding, angle encoding, and improved angle encoding. We further describe accompanying optimization that can be applied to angle and improved angle data encoding.
Finally, we demonstrate each method’s implementation and compare the efficacy of each method using quantum cost, quantum depth, and the number of required qubits.
Future work will explore how the use of these different encoding approaches affects the structure of various well-known QC algorithms.

\section{Background}
\subsection{Quantum Read-Only Memory}
\begin{figure}[htbp]
\centerline{\includegraphics{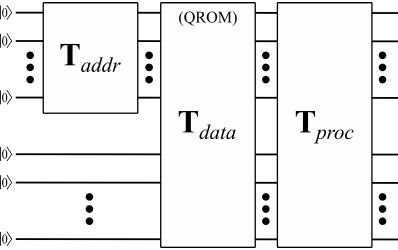}}
\caption{Generalized quantum algorithm with embedded QROM structure.}
\label{QROM_circs}
\end{figure}

A quantum read-only memory, or QROM, is a quantum state-generation circuit that evolves initialized qubits into a quantum representation of addressable, classical data.  Thus, a QROM is an addressable $2^n \times m$-qubit state-generation circuit wherein $n$ qubits serve as address values that select one or more of the $N=2^n$ data words to be generated, each with a wordsize of $m$ qubits.  Addressable QROMs are advantageous, because a superposition of address values can be used to generate a superposition of stored data values for a quantum algorithm to process.  Figure ~\ref{QROM_circs} illustrates a quantum algorithm's use of a QROM in the form of a generalized quantum circuit that consists of three distinct portions.  The leftmost portion, $\mathbf{T}_{addr}$, selects the addresses whose data is to be processed; the middle portion, $\mathbf{T}_{data}$, is the QROM state-generation circuit that is automatically synthesized by our compiler; and the rightmost portion, $\mathbf{T}_{proc}$, consists of the data-processing portion of the quantum algorithm.  Should the QROM need to be accessed again, after $\mathbf{T}_{proc}$ executes, the second portion, $\mathbf{T}_{data}$, can be appended to the rightmost portion, $\mathbf{T}_{proc}$, of Figure ~\ref{QROM_circs}.
By taking advantage of a Hilbert space that grows exponentially with the number of address qubits, this addressable QROM circuit requires $n=log(N)$ address qubits entangled with each of the $N$ data values~\cite{nielsen2002quantum}.

Reference~\cite{babbush2018encoding} introduces the concept of a QROM as described here with a sub-circuit termed ``unary iteration" that consists of control lines, address lines and accumulator output lines.
Reference~\cite{babbush2018encoding}'s basic structure is a series of multiple-controlled-NOT gates (MCX gates), where Pauli-$\mathbf{X}$ gates are applied as the target operation for the corresponding output qubits in what is similar to the approach we define here as ``basis encoding.''
Reference~\cite{babbush2018encoding} then performs an optimization referred to as ``sawtooth optimization,'' that allows cancellation of many Toffoli gates at the cost of introducing additional ancilla qubits as targets.
In Reference~\cite{phalak2022optimization}, an optimization analogous to the creation of classical pre-decoded signals, that is, the signals that make up the intermediary inputs of a wide-input gate, is used to optimize Reference~\cite{babbush2018encoding}'s QROM method.
Specifically, the optimization requires use of more qubits while reducing the exponential increase in depth that is a consequence of adding address control lines and MCX gates.

Most other related work involves the creation of QRAM, which we briefly discuss here to differentiate it from QROM.
QRAM uses a bifurcation graph to represent the RAM as a binary tree.
Using qutrits as nodes of the tree allows for reduction of active superposition paths, which improves fault tolerance~\cite{Giovannetti2008}.
QRAM has the same exponential circuit complexity as QROM, and includes extra hardware to make qubits writable from a classical source, but it does not require synthesizing data at the quantum gate level.
One version of QRAM, termed FF-QRAM, uses a series of classically-controlled NOT gates to compute and uncompute the address line inputs to retrieve a superposition of continuous rotations~\cite{park2019circuit}.

While some quantum algorithms---including some in QML---require the ability to write to QRAM, most do not, meaning a QROM is sufficient.
Furthermore, using a QRAM is more difficult than a QROM for at least two reasons.
First, generating a QRAM requires an assumption that classical-to-quantum mappings of the desired data are readily available, which is often not the case~\cite{Giovannetti2008}.
Second, the re-writable nature of a QRAM's qubits makes QRAM hardware more challenging to construct than the quantum gates required for QROM circuits~\cite{bertels2021quantum}.
Although there are some applications for which a QRAM is a more desirable choice---specifically, when new values need to be stored within the memory---the remainder of this paper focuses on QROM implementations.

Structures such as QROM are of particular importance since many algorithms benefit from a quantum speedup \textit{only} when multiple words of data are accessed and processed simultaneously through the use of superposition.
QROM methods are generally based upon representing memory values in a basis-encoded form, which is the most straightforward and intuitive approach, but which offers the least efficiency in terms of required numbers of qubits~\cite{Schuld2018,ventura2000quantum, phalak2022optimization}.

\subsection{Data Encoding}
In any field requiring the processing of significant amounts of data, the manner in which the data is encoded matters.
For example, in the field of QML, developing different forms of encoding data is important, because machine learning depends upon input data that often requires situation-dependent pre-processing\cite{sierra2020tensorflow}.
Therefore, a number of encoding methods have been developed for QML, including amplitude encoding, angle encoding, Q-sample encoding, and hardware-specific encoding approaches~\cite{Schuld2018,schuld2021effect,weigold2020data, ventura2000quantum}.
Although we have results for several different types of data encoding approaches, given space constraints in this paper, we focus only on basis and angle encoding, as well as a novel modification thereon that we term ``improved angle encoding.''

Basis and angle encoding can be conveniently defined in terms of parameterized qubit models. A qubit, $\ket{\Psi}$, can be mathematically modeled as $\ket{\Psi}=\alpha\ket{0}+\beta\ket{1}$, wherein the parameters $\{\alpha, \beta\}\in\mathbb{C}$ are probability amplitudes that are so-named because they can be used to calculate the probability of the outcome of a qubit measurement with respect to some measurement basis. A qubit may also be parameterized within the spherical coordinate system as $\ket{\Psi}=cos\frac{\theta}{2}\ket{0}+e^{i\phi}sin\frac{\theta}{2}\ket{1}$, where $\theta$ represents longitudinal position on the Bloch sphere, and $\phi$ is the phase difference amongst the probability amplitudes, or the ``phase'' of the qubit.

\begin{definition} \label{basis-enc}
\textbf{Basis encoding} restricts $\{\alpha, \beta\}$ to either the set $\{1,0\}$ or $\{0,1\}$.  In this case, a qubit may represent a single bit of information, either zero (0) or one (1).  Thus, a value comprising $m$ bits requires $m$ qubits. The mapping can be denoted in binary as $(b_{m-1}b_{m-2}, \cdots, b_1b_0)\rightarrow \ket{b_{m-1}b_{m-2}, \cdots, b_1b_0}$, where $b_i \in \mathbb{Z}_2$. \hfill $\Box$ 
\end{definition} 

\begin{definition} \label{angle-enc}
\textbf{Angle encoding} represents classical values using the parameterized qubit angles $\{\theta, \phi\}$.  Thus, two $m$-bit values can be represented by a single qubit.  Angle encoding requires the memory image to be interpreted, or possibly pre-normalized, such that all memory values are within the range $[0,2\pi)$. The mapping can be shown as $ x \in [0, 2\pi)\rightarrow cos(x_j)\ket{0}+e^{ix_{j+1}} sin(x_j)\ket{1}$ where two classical data values are represented as $x_j$ and $x_{j+1}$ respectively.
\hfill $\Box$
\end{definition} 

\begin{definition} \label{imp-angle-enc}
\textbf{Improved angle encoding} also represents classical values using the parameterized qubit angles $\{\theta, \phi\}$.  In this case, the angle $\theta$ represents the significand, $S$, of an $m$-bit value and the phase angle $\phi$ represents an integer exponent, $E$.  Thus, a single $m$-bit value, $V$, can be represented by a single qubit as $V=S\times2^{E}$.   The mapping can be shown as $ S, E \in [0, 2\pi)\rightarrow cos(S)\ket{0}+e^{iE}sin(S)\ket{1}$. \hfill $\Box$
\end{definition}

Given these definitions of data encoding types, we can define a mathematical model of a QROM as a set comprising two $N$-dimensional vectors, $\{\vec{a}, \vec{x}\}$ where the $j^\text{th}$ components of each vector comprise the $n$-bit address field, $a_j$ and the $m$-bit data word $x_j$.  Therefore, in terms of information content, the QROM comprises $N=2^n\times m$ bits of information, although the representation of the information varies depending upon the data encoding method that is employed.  Table \ref{Encoding_Summary} summarizes the mathematical models of the encoding methods and QROM types that are the focus of this paper.

\begin{table*}[ht]
\caption{Summary of Data Encodings\label{Encoding_Summary}}
\begin{center}
\begin{tabular}{|c|c|c|}\hline
\hline
Encoding Type & Classical Data & Quantum State \\
\hline
Basis Encoding & $x \in \mathbb{Z}_2^m$, $x=(b_{m-1}, b_{m-2}, \cdots,b_0)$, $b_j \in \mathbb{Z}_2$ & $\ket{x} = \ket{b_{m-1} b_{m-2} \cdots b_0}$ \\

Angle Encoding &  $x \in [0, 2\pi)$ & $\otimes_{j=0}^{(N-2)/2} cos(x_{2j+1})\ket{0}+e^{ix_{2j}}sin(x_{2j+1})\ket{1} $\\

Basis QROM & $\{\vec{a},\vec{x}\}, a_j \in \mathbb{Z}_2^n, x_j \in \mathbb{Z}_2^m$ & $\ket{\psi_x} = \sum_{j=0}^{N-1}{\ket{a_j}\ket{x_j}}$\\

Angle QROM & $\{\vec{a},\vec{x}\}$, $a_j\in \mathbb{Z}_2^n, x_j \in [0, 2\pi)$ & $\sum_{j=0}^{(N-2)/2} \ket{a_j} (cos(x_{2j+1})\ket{0}+e^{ix_{2j}}sin(x_{2j+1})\ket{1}) $\\

Improved Angle QROM & $\{\vec{a},\vec{x}\}$, $a_j\in \mathbb{Z}_2^n, S_j \in [0, 2\pi), E_j \in [0, 2\pi)$ & $\sum_{j=0}^{N-1} \ket{a_j} (cos(S_j)\ket{0}+e^{i E_j}sin(S_j)\ket{1} )$\\

\hline
\hline
\end{tabular}
\end{center}
\end{table*}

\subsection{Applications and use of QROM}
QROMs have widespread applicability, particularly in the emerging area of QML where data set training and operational data input are required.  To motivate the use of QROM state generation circuits, we summarize examples of QC algorithms that process data by employing different data encoding approaches.

State vector machines (SVM) implemented on quantum computers can use an angle-encoded QROM.
Specifically, the $M$-dimensional data vector $\vec{x} \in \mathbb{R}$  is stored in $|\chi\rangle=\frac{1}{\sqrt{\sum_{j=0}^{N-1}\left|\vec{x}^{(j)}\right|^2}} \sum_{j=0}^{N-1} \sum_{k=0}^{M-1} x_k^{(j)}|k\rangle|j\rangle$, where $N$ is the number of vectors, or the data set size, $k$ is the address, and $j$ is an additional entangled ordered index~\cite{rebentrost2014quantum}.
This SVM data representation is very similar to the angle-encoded QROM described here but with an additionally entangled index register.

Additionally, QROMs need not be used solely as inputs to a model; they can be used as part of a model itself. One such example is quantum weightless neural networks (WNNs) that use QROMs as subcircuits.
Unlike many neural network architectures, WNNs use table lookups, instead of a sum of multiplied operations.
Reference~\cite{da2012classical} describes such a model, and our automated quantum memory compiler can serve as a tool that creates each individual table as a QROM node in the WNN.

Furthermore, QROMs can be applied in the implementation of machine learning kernel calculations.
For example, Reference~\cite{schuld2021supervised} discusses the benefits of different data encoding approaches for kernels that support a variety of machine learning algorithms, including SVMs and clustering methods.
Additionally, Reference~\cite{sinha2022quantum} implements a QROM subcircuit for creating an encoding-dependent, parallelized quantum kernel in the higher-dimensional, or ``qudit,'' case.

\subsection{Data Encoding Approaches}
Beyond the direct application of QROM circuits in QML, Reference~\cite{schuld2021effect} considers how the choice of encoding method impacts quantum neural network (QNN) training wherein QNNs are sometimes considered a subset of the more general class of variational quantum algorithms.
Similarly, Reference~\cite{sierra2020tensorflow} provides experimental results regarding the implementation of different encoding approaches for data representation in the Python programming language's \texttt{TensorFlow} package, and it examines the effect of data encoding on the performance of QML algorithms.

Results such as those in Reference~\cite{sierra2020tensorflow} and Reference~\cite{schuld2021effect} illustrate that different encoding approaches affect the fault tolerance, performance, and ``learnability" of QNN, as is further discussed in Reference~\cite{larose2020robust}.
Our automated QROM compilation tool supports further evaluation of encoding types and their effects on circuit performance, because it allows for using multiple data encoding methods on the same dataset, which can then be compared for applications such as Q-means clustering~\cite{kerenidis2019q} and weightless networks~\cite{da2012classical}.

\subsection{Memory Image Pre-processing}
Because the square of probability amplitude magnitudes and parametric angular values must fall within specified ranges of $[0,1]$ and $[0,2\pi)$ respectively, the data encoding process requires special rules for interpreting or normalizing memory image values.  From a theoretical point of view, infinite precision is available in an angle-encoded QROM, since $(\theta, \phi) \in \mathbb{R}$. However, our use of memory image files containing discrete and limited resolution values as input to the circuit synthesizer necessarily restricts the precision with which memory values are represented.  It is noted that the advantage of infinite precision can be exploited within intermediate calculations in a quantum circuit, but further examination of this topic is beyond the scope of this paper.

There are many approaches for normalization; the most naive method is prone to both inefficiency and inaccuracy, so we discuss it alongside more efficient and accurate alternatives.

Before turning to those details, however, we note that one benefit of basis encoding is that no memory image normalization is required, thus avoiding potential loss of precision.
Unfortunately, this precision comes with a cost: basis-encoded QROMs use qubits inefficiently, requiring $2^n \times m$ qubits to implement a QROM of the same size.
Thus, as the size of the memory increases, basis-encoded QROMs require significant numbers of qubits that are often unavailable on contemporary machines.

\subsubsection{Angle Encoding Pre-processing}
There are a variety of normalization approaches that can be applied to an angle-encoded QROM.
In one approach, each of the two values stored within a single qubit are normalized to be within $[0, 2\pi)$ by creating a function that finds the maximum value in a memory image, $V_{max}$, computes a normalization factor, $f_{norm}={2\pi}/V_{max}$, and then multiplies each value to be encoded by $f_{norm}$.
Separate $V_{max}$ and $f_{norm}$ values are required for the set of memory words stored in $\theta$ and for those stored in $\phi$.
The multiplication operation must be applied to each memory value using the appropriate $f_{norm}$ normalization factor.

A more efficient approach uses an encoding method for angles $\{\theta, \phi\}$ wherein each memory value is assumed to be a positive, fixed-point value, with the fixed-point position determining the range into which the data will fall.
This approach interprets each datum as having the form $0.b_{-1}b_{-2} \cdots b_{-m+3}b_{-m+2}$ where $b_j \in \mathbb{Z}_2$, allowing for values in the range $[0, 1)$, which is within $[0, 2\pi)$.
This is an interpretative approach that avoids computation of $V_{max}$ and performance of $N=2^n$ multiplication operations.
However, memory word resolution is sacrificed since the full available range of $[0, 2\pi)$ is not used, given that all memory values are interpreted as within the range $[0,1)$.
If desired, each memory value can be scaled by multiplying each memory word by $f_{norm}=2\pi$ to enable utilization of the full range, which avoids the computation of $V_{max}$ values, but requires $N=2^n$ multiplication operations.

An improved variant of this fixed-point interpretative method allows for utilizing a larger portion of the available $[0, 2\pi)$ range, thus increasing the resolution of each value generated by the angle-encoded QROM circuit.
Assuming that the fixed-point value has the form $b_1b_0.b_{-1}b_{-2} \cdots b_{-m+3}b_{-m+2}$, all possible memory word values fall within the range $[0, 4)$.
Although this method involves some loss of precision, since it does not use the entire available interval of $[0, 2\pi)$ (because $[0, 4)\subset[0, 2\pi)$), it is a more efficient pre-processing method requiring no normalization and offering increased memory word resolution when compared to the previous method that condenses all values into the smaller $[0, 1)$ range.

It is important to note that our definition of angle encoding precludes the use of sign bits, since memory words are stored in real-valued parametric qubit angles, $(\theta, \phi) \in \mathbb{R}$.  However, signed values can be represented using angle encoding by either normalizing or interpreting memory data values to fall within the intervals $[-\pi,\pi)$ or  $(-\pi,\pi]$ rather than $[0, 2\pi)$ or $(0, 2\pi]$.  

\subsubsection{Improved Angle Encoding Pre-processing}\label{IAE_Normalization}
If a memory image contains both very large and very small magnitude values, we characterize it as having a large ``dynamic range.''
Normalization of memory images with a high dynamic range can result in a significant loss of precision for the smaller magnitude values.
Specifically, the multiplication of smaller memory words by $f_{norm}$ can result in a fixed point value with a large number of leading zeros that, in turn, can cause the loss of significant bits due to a finite word size.
While this issue could be partially addressed through the use of floating-point arithmetic during the normalization process, additional issues can arise due to the conversion process among fixed- and floating-point representations notwithstanding the significant performance penalties arising from the use of floating-point versus fixed-point arithmetic.
This class of resolution errors worsens as the memory image dynamic range increases and as the memory wordsize $m$ decreases.

Improved angle encoding is intended to counteract this loss of precision by representing memory values in a manner that is akin to floating-point representation in conventional digital systems.
Specifically, each qubit represents the $j^\text{th}$ memory word value in the form $S_j\times2^{E_j}$, where $S_j$ is the ``significand'' and $E_j$ is the corresponding ``exponent.''

The first step in improved angle QROM pre-processing requires the computation of the number of leading zeros for each word in the memory image, denoted as $z_j$ for the $j^\text{th}$ memory word.
The maximum number of leading zeros for any given word, $z_{max}=max\{z_j\}$, defines the upper bound of possible exponent values, $E_j \in \{0, 1, \cdots, z_{max}\}=\mathbb{Z}_{z_{max}}^+$.
The significand, $S$, contains values equivalent to a fixed-point interpretation of the $m$-bit memory words that are left-shifted by $z_j+2$ bits and right-padded with $m-(z_j+2)$ zeros.
As with angle encoding, each left-shifted memory word is interpreted according to the position of the fixed-point that is assumed to fall between the second and third bits from the leftmost side of the shifted bit string.
Specifically, the significand, $S_j$, is interpreted as the value $S_j \in [0,4)$ since its left-shifted form is $S_j=(b_1b_0.b_{-1}b_{-2} \cdots b_{m-z_j}00\cdots0$) where $b_1=1$ due to the shifting operation. 
In general, improved angle encoding can take advantage of any of the above mentioned options for angle encoding normalization or interpretation.

The exponent, $E$, is stored in the qubit phase angle and must likewise be normalized or interpreted to fall within the range $[0, 2\pi)$.
A simple normalization method is to apply a multiplicative factor equivalent to $\frac{z_{max}}{2\pi}$ to the $z_j$-values for each $j^\text{th}$ word.
It is noted that, because $E \in \mathbb{Z}_{z_{max}}^+$ is an integer, the real-valued interval $[0, 2\pi)$ could be partitioned into $z_max$ sub-intervals, and a given valuation of $E$ could be assigned to any value within the appropriate sub-interval.
This option can be exploited to enable optimizations in the synthesized QROM circuit; however, it would likewise complicate the process of recreating the original memory value during a QROM read operation.
One possibility is to implement a ``round-to-nearest integer'' rule to recreate the integer-valued exponent, $E$.
An added advantage of such a round-to-nearest rule is that small errors or inaccuracies induced in the $E$ value due to the use of practical, non-ideal quantum gates would have some degree of self-correction present.

Yet another method that further increases precision without requiring normalization is to use a ``hidden bit," as does the IEEE 754 standard~\cite{ieee2019}. 
Due to the left-shift operation in determining the significand, the most significant bit (MSb) is always set to one (1), $b_1=1$.
If a hidden bit is implemented, one extra left shift could be accomplished to enable an extra least significant bit to be included in the significand.
In this case, the MSb, $b_1$, could be either zero or one with the assumption that an inherent MSb is present and equal to one, but is not included in the stored value within the parametric qubit angle.
Thus, use of a hidden bit offers one more bit of increased precision by omitting the leading one (1) bit of the left-shifted significand.

\section{Automated QROM Compilation}
The techniques for compiling a QROM circuit given a conventional memory image are implemented in an automated quantum computer compilation tool, MustangQ.
MustangQ targets the ``gate model'' paradigm of quantum computation and produces output in the form of OpenQASM that can be transpiled for execution on different QC~\cite{smith2019QCcompile}.
Due to the similarities in compilation of quantum circuits for execution on QC and the design of application-specific quantum integrated circuits (QIC), MustangQ is designed to be both a QC compiler and a quantum design automation (QDA) tool for QIC~\cite{smith2019quantum}.
This section focuses on the aspects of MustangQ that are specifically relevant to memory compilation.

\subsection{MustangQ for QROM Compilation}
MustangQ is intended to serve as a general tool for quantum circuit synthesis, optimization, verification, and visualization that targets both QC and QIC.
Thus, MustangQ comprises a wide variety of methods and techniques; however, only those features dedicated to the automatic synthesis of QROMs are described in detail.

The initial specification of the QROM is provided to MustangQ in the form of a classical memory image in \texttt{.pla} file format \cite{rudell1986multiple}.
Although the \texttt{.pla} format was originally intended to describe a digital electronic circuit in the form of a ``programmable logic array,'' we determined that it is particularly useful for describing memory images since it supports arbitrary-sized address fields $n$ and memory word sizes $m$.
It also provides a degree of compression.
One compression method is due to the use of ``output don't cares'' that combine same-valued bits in memory words into a single entry.
Another compression method is to combine same-valued address bits into a single entry through the use of ``input don't cares.''

The compiler is capable of generating circuits with fifteen (15) different quantum gates, six (6) of which are single-qubit and eleven (11) of which are multi-qubit gates.
A particular QROM corresponding to an input \texttt{.pla} file is output in the form of an OpenQASM file that can be imported into the IBM Qiskit environment for use in quantum algorithms that execute on the IBM QC. 

The QROM synthesis functionality incorporates various quantum circuit optimizations in addition to mapping the irreversible memory specification of a \texttt{.pla} file into a corresponding reversible QROM circuit.
Examples include the removal of multiple single-qubit Pauli-$\mathbf{X}$ gates that equate to an identity function and the decomposition of multi-control gates into single-control gates.
A total of five (5) different optimization transformations are included in the QROM synthesis functionality of MustangQ.

In addition to the mapping and optimization functions, the three different data encoding approaches previously described allow flexibility in the form of the resulting QROM memory circuit.
The particular optimizations applied to the QROM during synthesis vary depending upon the user-selected data encoding method; thus, they are described in more detail in the following subsections.

\subsubsection{Memory Images in the \texttt{.pla} Format}
The \texttt{.pla} file format was originally developed to represent a switching function with $n$ bits comprising the domain values, $m$ bits comprising the corresponding range values, and each line storing one or more valuations of the function in a ``sum-of-products'' (SOP) form.
As is well-known in the switching theory community, a memory circuit can be viewed as an SOP switching function wherein the minterms are represented by address values and the $m$-bit function value is represented by the memory data word.
We adopt this interpretation of a \texttt{.pla} file for the purposes of the QROM compiler tool, meaning a particular \texttt{.pla} file is interpreted as a memory image for an addressable memory that has an $n$-bit address field and an $m$-bit word size.
Thus, each \texttt{.pla} file represents a $2^{n\times m}$ bit memory, since it could store as many as $2^{n \times m}$ bits or $2^{(n-3)\times m}$ bytes.


\begin{figure}[htbp]
\centerline{\includegraphics{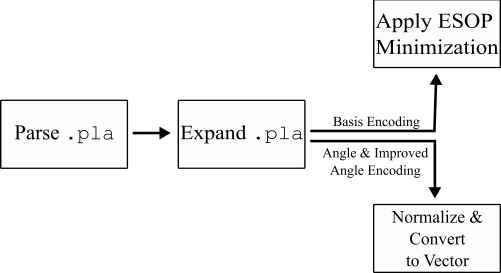}}
\caption{Preprocessing for classical memory images stored as 
\texttt{.pla} files.}
\label{Preprocessing_Summary}
\end{figure}

\subsubsection{Encoding Pre-processing}
Although all encoding approaches share a standardized input format---here, \texttt{.pla} files---the pre-processing  operations of the memory image values vary depending upon the desired QROM encoding.
Figure ~\ref{Preprocessing_Summary} summarizes the pre-processing flow for each of the three data encoding types.
First, the content of a \texttt{.pla} file is parsed into a custom data structure that enables a more efficient way of pre-processing the memory image and avoids excessive file input and output.
Second, the represented memory image is expanded, which means that the function is represented without the size reduction allowed by input don't cares.
Although this significantly increases the storage required for the pre-processed memory image, it allows straightforward access to each address/data pair, which is important in later steps of the pre-processing.
We note that the expansion process could likely be avoided with more sophisticated pre-processing algorithms; however, for the proof-of-feasibility approach in this paper, we apply the simpler technique of expansion.
Additionally, expansion involves inserting any missing address/data pairs explicitly that may have been omitted in the original \texttt{.pla} file.
This is important, because the memory image needs to be completely specified; however, for the original intent of \texttt{.pla} files---representing switching functions---it is often the case that such functions have unspecified input/output pairs.
For the results in this paper, we set all missing memory words to zero.
A future optimization would allow the synthesis tool to choose unspecified memory word assignments according to the optimization criteria.

The third step differs depending upon the encoding type applied. For basis encoding, exclusive-or-sum-of-products (ESOP) minimization is applied to the memory image as was first described in~\cite{fazel2007esop}.
While any ESOP minimization tool could be used, we use EXORCISM-4 for this step~\cite{mishchenko2001fast}.
For either angle or improved angle encoding, data must be converted into a real number.
Thus, each $m$-bit memory word is interpreted as a radix-10 value in the range $[0, 1)$ or $[0,4)$.
For example if $m=4$ and $x_j=1111$, $x_j$ is interpreted as a fixed-point value of the form $(0.1111)_2$ or as $(11.11)_2$, corresponding to $(0.9375)_{10} \in [0,1)$, or $(3.75)_{10}\in [0,4)$, respectively.
In our implementation, we create an intermediate array, $\vec{x}$ with each array index corresponding to the address values $\vec{a}$.
Again, it is noted that this exponentially large array could be avoided through the use of alternative data structures and representations, but it is implemented and described here for ease of understanding.

\subsection{Synthesizing the QROM}
\subsubsection{Basis-Encoded QROM}
A basis-encoded QROM can be described in Dirac notation as provided in Table~\ref{Encoding_Summary}.
Basis encoding represents memory words as computational basis vectors in a Hilbert vector space of dimension $m$, the memory word size, denoted as $\mathbb{H}^{m}$.
Specifically, basis encoding requires the use of one qubit per bit comprising a memory word, with the overall QROM requiring $n+m$ qubits.

\begin{figure}[htbp]
\centerline{\includegraphics{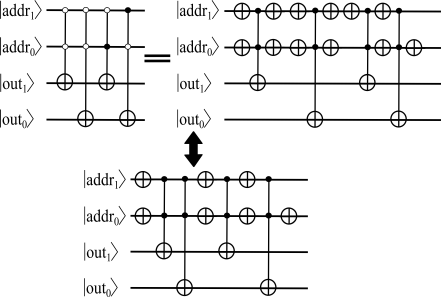}}
\caption{A sample QROM with basis encoding. The first line illustrates the unoptimized QROM, while the second illustrates an optimized version with lower quantum cost and quantum depth.}
\label{Basis_QROM}
\end{figure}

Details of the synthesis method for basis encoding are described in~\cite{fazel2007esop} and are briefly summarized here for completeness.
The QROM consists of $n+m$ instantiated qubits where $n$ qubits are used to specify the address, $a_j$, and $m$ qubits provide the memory word value, $x_j$.
The $m$ qubits corresponding to the memory word are initialized to $\ket{0}$.
Each memory word in the image file is processed by applying a control point to the address qubit corresponding to a one-valued address bit and a so-called ``negative control'' to a zero-valued address bit.
Negative controls are activated with a qubit value of $\ket{0}$ and can be realized as a normal control point that is surrounded by Pauli-$\mathbf{X}$ gates as shown in the topmost portion of Figure~\ref{Basis_QROM}.
The control qubits activate a Pauli-$\mathbf{X}$ target; thus, the overall gate is a generalized Toffoli gate.
After the cascade of generalized Toffoli gates is generated, they are decomposed into subcircuits of single- and two-qubit gates and then further transformed according to the optimization criteria as shown in the bottom-most portion of Figure~\ref{Basis_QROM}.
Using a simple cost function wherein each control and single qubit operation have unity cost, the resulting optimized QROM of Figure~\ref{Basis_QROM} results in a 20\% cost reduction.

Basis encoding is the most straightforward approach and is the assumed method of data representation for many quantum algorithms.
For example, the Quantum Fourier Transform (QFT) maps input data stored using computational basis vectors into an alternative basis, known as the Fourier basis~\cite{nielsen2002quantum}.

\subsubsection{Angle-Encoded QROM}
Angle-encoded QROM can be described in Dirac notation as provided in Table~\ref{Encoding_Summary}.
During synthesis, multi-controlled rotation gates, $\mathbf{R}_x$ and $\mathbf{R}_z$, are generated with controls on the address qubits and with the target rotation applied to the single qubit comprising the memory word value.
The controlled-$\mathbf{R}_x$ are used to specify the values stored within the probability amplitudes, and the controlled-$\mathbf{R}_z$ are used for the values stored within the phase angle.
Thus, a chief advantage of angle encoding is the use of a single quantum state---regardless of the initial memory word size, $m$---versus the use of $m$ quantum states for the memory words.
As in the case of basis-encoded QROM, a superposition of specified address qubit values can be used to retrieve corresponding superimposed values of the memory words.

Angle encoding can be used to store two memory words within a single qubit's quantum state by making use of both parametric qubit angles $\{\theta, \phi \}$; we refer to this case as ``dense angle'' encoding.
Dense angle encoding requires that two different methods of data word access be implemented; one method accesses the data within the probability amplitude, and the other requires a phase detection process.
As described in Table~\ref{Encoding_Summary}, memory words with odd addresses are stored in the probability amplitudes, and those with even addresses are stored in the qubit phase.
Thus, the qubit that represents the LSb of the address can be used to select the appropriate form of memory word access.
One disadvantage of dense angle encoding is that measurement of a word at an odd (or even) address will destroy the memory value in the corresponding even (or odd) address. 

In the results section of this paper, we implemented the use of probability amplitudes only for angle-encoded memory word storage.
In this case, the qubit phase value is irrelevant and may be exploited as a degree of freedom in the QROM optimization algorithms.
Depending upon the QROM use-case, choosing to use dense angle encoding might be preferable to using the qubit probability amplitudes (or phases) only.

\subsubsection{Improved Angle-Encoded QROM}
Improved angle-encoded QROM has a mathematical structure very similar to that of angle encoding and is given in Dirac notation in Table~\ref{Encoding_Summary}.
Unlike the case of dense angle encoding, a single memory word is encoded within a single qubit state.
Specifically, a memory value $V$ is represented as a significand $S$ using the probability amplitude, $\theta$, with the corresponding exponent value $E$ encoded within the phase angle, $\phi$.
So, $V=S \times 2^E$.

\begin{figure}[htbp]
\centerline{\includegraphics[scale=2]{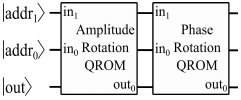}}
\caption{The structure of improved angle-encoded QROM.}
\label{ImprovedAngleFig}
\end{figure}

Because the critical difference between improved angle encoding and angle encoding is the data storage and pre-processing procedure, the synthesis process is much the same as that described for angle-encoded QROM.
As illustrated in Figure \ref{ImprovedAngleFig}, improved angle encoding involves the synthesis of two stages of circuitry, with the first performing transformations that affect the qubit probability amplitude and the second transforming the qubit phase.
The first stage makes use of multi-controlled rotation gates about the $x$-axis, $\mathbf{R}_x$, and the second stage makes use of multi-controlled rotation gates about the $z$-axis, $\mathbf{R}_z$.
Although we conceptually describe and visualize the QROM as comprising two serial cascades, many of the operations can be re-ordered, allowing for additional degrees-of-freedom during QROM optimization.

One example for which the use of improved angle encoding is preferable to angle encoding is the case in which two retrieved QROM values are multiplied.
If one of the values is very low in magnitude and the other very large, underflow could occur, causing the resulting product value to be zero.
Conversely, the use of improved angle encoding would allow for a non-zero product value to be computed.
This example illustrates the advantages of improved angle encoding, since it effectively increases the dynamic range of values that can be represented and computed within a quantum circuit.

\subsection{Gray-Code Optimization}\label{Gray_Code_Optimization}
For both angle and improved angle QROMs, the most-oft-applied gates are multiple-control-rotation gates, which can be abbreviated as $\mathbf{cR}_x$, $\mathbf{cR}_y$, $\mathbf{cR}_z$, depending upon whether the target rotation is about the $x$, $y$, or $z$ axes.
As described in~\cite{nielsen2002quantum}, $\mathbf{cR}_x$ gates can be very expensive in terms of quantum cost and quantum depth.
There is thus cause to employ an optimization technique that reduces the size of consecutive multiple controlled rotation gates in what we term ``gray code'' optimization.
The gray code optimization is motivated and extended from the results of ~\cite{daskin2014universal}.

In the QROM circuits described here, the address qubit lines are represented by unique $|j\rangle$ basis address values and the multi-controlled gate controls are synthesized with control points to activate the rotation at the corresponding address lines.
For every unique set of controls, there is a corresponding rotational value which represents the angle for the data associated with the address specified by the controls.
We note that, in theory, the address values could also be encoded within the probability amplitudes and/or qubit phase values; however, we reserve this case for future research.

A key observation is that the order in which transformations are applied to generate different memory words is generally not of importance, and that the naive method of generating the memory words in the order of ascending address values is not required.
Therefore, we can re-arrange the order in which different memory words are generated in the QROM, and in fact, we choose an order in which a minimal number of address bits change: a gray code order that results in decreased QROM quantum cost.

The gray code optimization converts $2^n$ $\mathbf{cR}_x$ gates, each with $n$ address lines, to a series of $2^n$ single-controlled $\mathbf{Z}$ and single-controlled rotations about the $x$-axis ($\mathbf{R}_x$) gates.
This works because $\mathbf{cR}_x$ gates can be decomposed into a series of single-qubit $\mathbf{R}_x$ gates and single-controlled $\mathbf{R}_z$ gates, where the single-controlled $\mathbf{R}_z$ gates effectively introduce cancellation of some of the evolved values due to previous $\mathbf{R}_x$ operations.

The rotation values for the optimized $\mathbf{cR}_x$ gates are computed using the results of~\cite{daskin2014universal}, where the original vector of $x$-axis rotation values applied to the $\mathbf{cR}_x$ targets is expressed in a vector $\vec{\rho}$ and is linearly transformed using a Hadamard matrix with permuted column vectors in the form of a gray code ordering, resulting in the transformed vector of rotations $\vec{\rho'}$.

\begin{figure}[htbp]
\centerline{\includegraphics[scale=2]{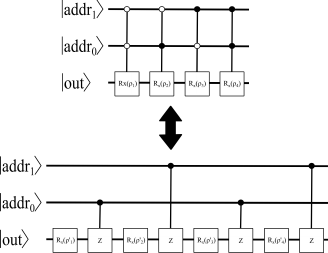}}
\caption{An angle-encoded QROM before and after gray-code optimization.}
\label{GraycodeOptFig}
\end{figure}

Figure~\ref{GraycodeOptFig} shows the originally implemented circuit with $\mathbf{cR}_x$ gates using angles in $\vec{\rho}$ being replaced with controlled-$\mathbf{Z}$ gates and single qubit $\mathbf{R}_x$ rotations using the vector $\vec{\rho'}$.
The $\mathbf{R}_x$ gates are interleaved between single-controlled $\mathbf{Z}$ gates.
The $\mathbf{Z}$ control locations follow a pattern similar to the depths of all nodes during an in-order transversal of a perfect binary tree.
We can obtain the control locations without construction of the binary tree itself by recursively subtracting one (1) from the number of address lines of the left subtree until it becomes the lowest control location, obtaining the current control location, and then repeating the same recursion on the right subtree.
This gray-code optimization reduces the number of gates, because a certain portion of the control qubits can be removed; thus, some of the rotation gates can be reused for other data values without changing the data stored. 

\section{Experimental Results}\label{Results_Section}
Quantum circuits can be evaluated using a variety of metrics, including gate count, qubit count, quantum cost, and quantum depth~\cite{smith2019quantum}.
Quantum gate count and qubit count are self-explanatory.
However, quantum cost and quantum depth are less standardized metrics~\cite{maslov2005comparison}.
Quantum cost is proportional to the number of gates in a circuit, but it weights different gates according to their properties~\cite{maslov2005comparison}.
We use a simple formulation in which the cost of a gate is equivalent to the number of qubits on which it acts, including both control and target qubits.
Quantum depth can be used to refer to varying quantities; we use the length of the critical path as computed by Qiskit~\cite{qiskitQuantumcircuit}.

\begin{figure}[htbp]
\centerline{\includegraphics[width=0.5\textwidth]{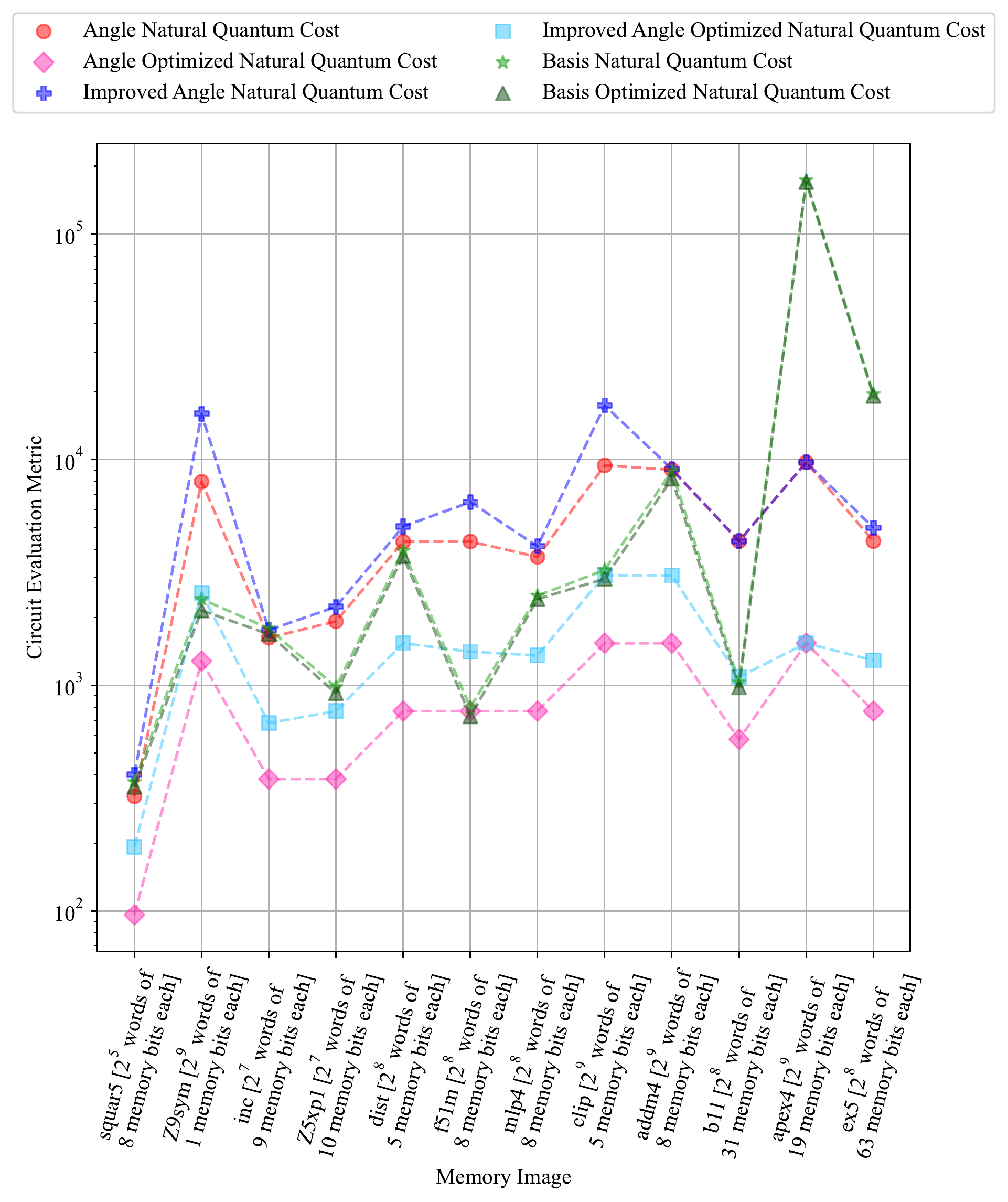}}
\caption{Quantum cost for six encoding/optimization situations using a ``natural'' gate set for each encoding. The points with red circles are for angle encoding, the points with pink diamonds are for gray-code-optimized angle encoding, the points with dark blue pluses are for improved angle encoding, the points with light blue squares are for gray-code-optimized improved angle encoding, the points with light green stars are for basis encoding, and the points with dark green triangles are for double-NOT-removal optimized basis encoding.}
\label{QuantumCost}
\end{figure}

\begin{figure}[htbp]
\centerline{\includegraphics[width=0.5\textwidth]{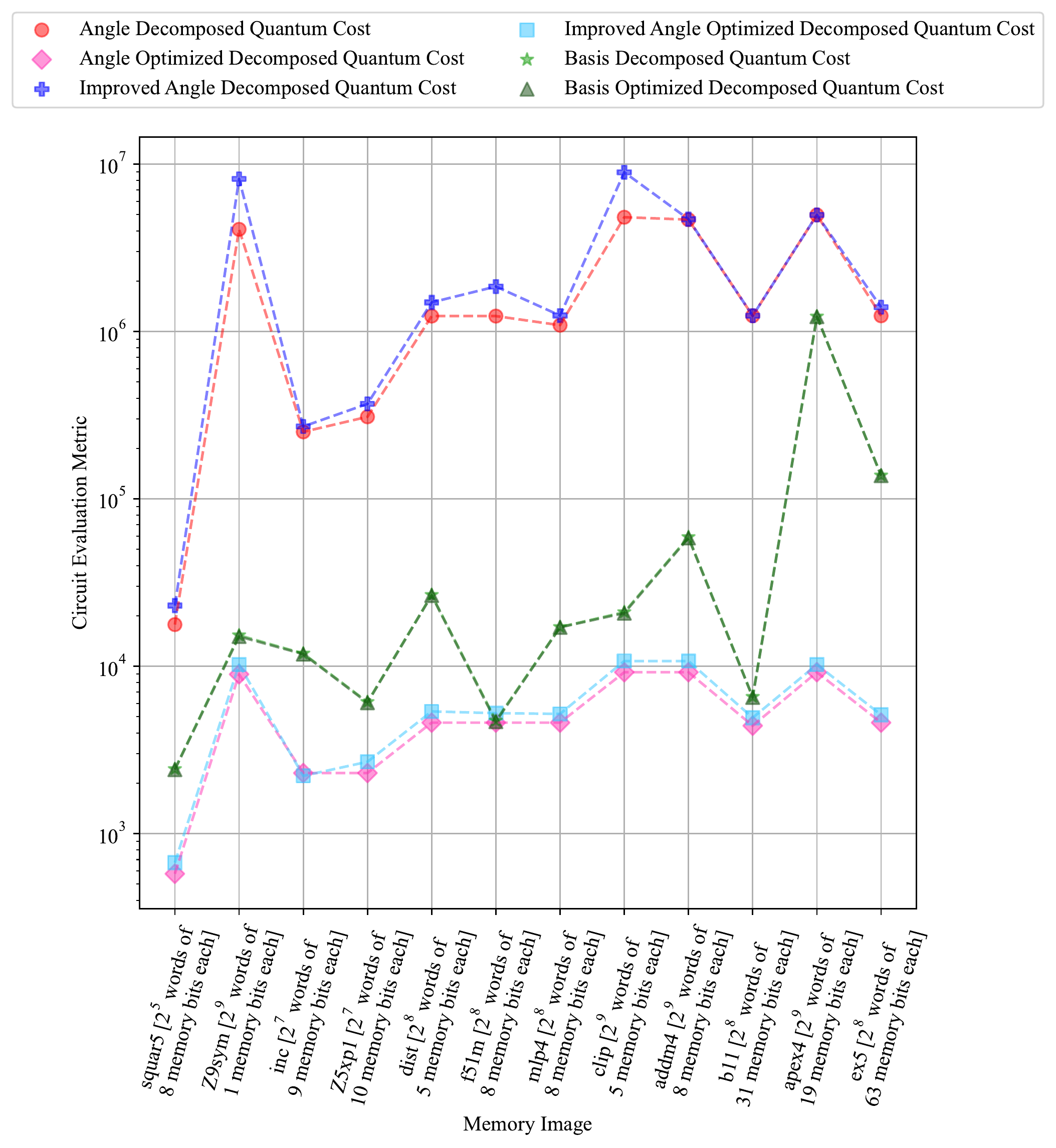}}
\caption{Quantum cost for six encoding/optimization situations using a uniform gatset for all encodings. The colors and lines are as described in Figure~\ref{QuantumCost}.}
\label{DecomposedQuantumCost}
\end{figure}

\begin{figure}[htbp]
\centerline{\includegraphics[width=0.5\textwidth]{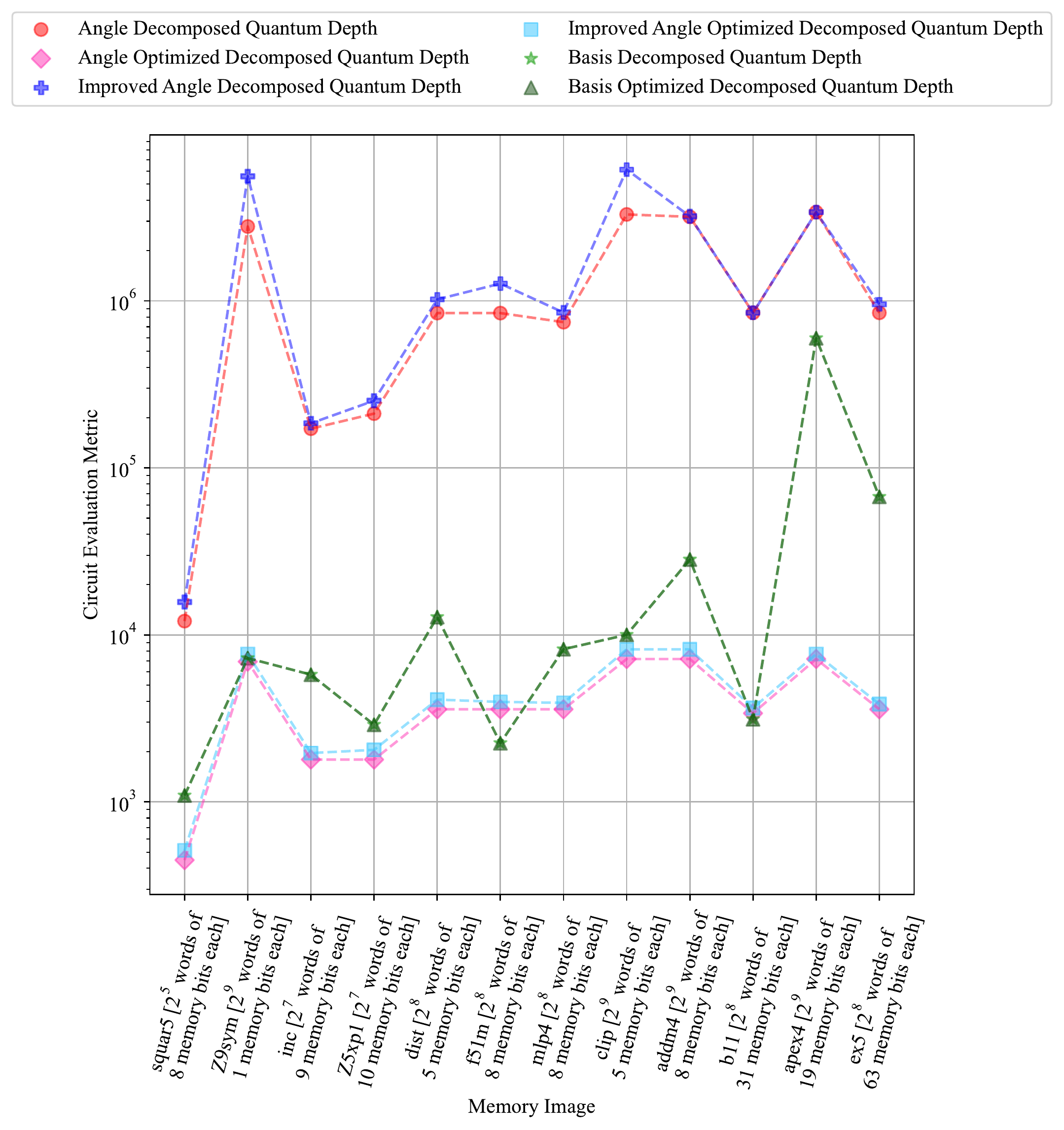}}
\caption{Quantum depth for six encoding/optimization situations. The colors and lines are as described in Figure~\ref{QuantumCost}.}
\label{QuantumDepth}
\end{figure}

\begin{figure}[htbp]
\centerline{\includegraphics[height=0.4\textheight]{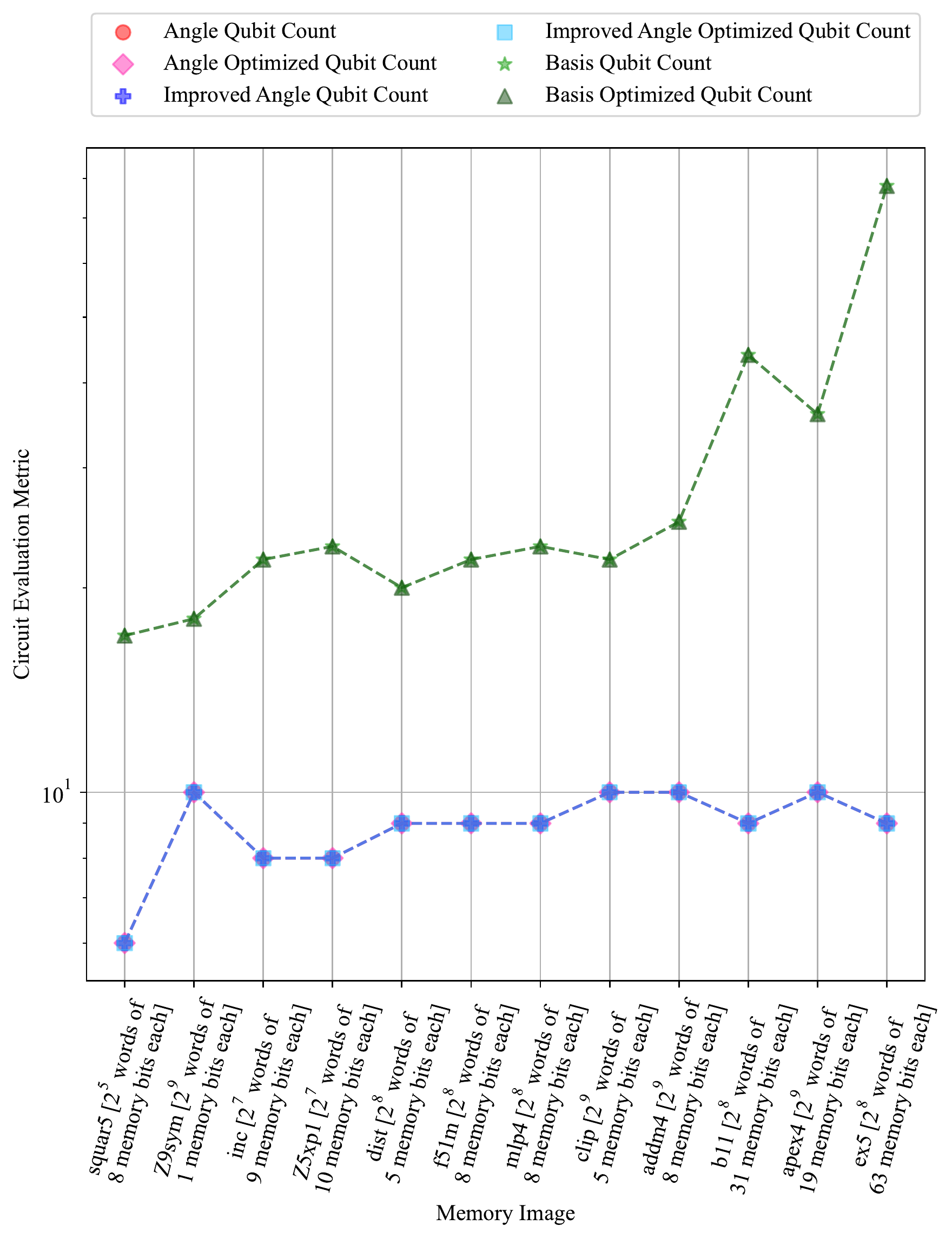}}
\caption{Qubit count for six encoding/optimization situations. The colors and lines are as described in Figure~\ref{QuantumCost}. Note that only two sets of points are visible because the optimizations applied do not reduce the number of qubits, and both angle encoding and improved angle encoding contain the same number of qubits. So, the additional sets of points are overlapped by the visible sets.}
\label{Qubits}
\end{figure}

Figures ~\ref{QuantumCost}, ~\ref{DecomposedQuantumCost}, ~\ref{QuantumDepth}, and ~\ref{Qubits} present statistics for a series of memory images specified as \texttt{.pla} files and synthesized using each encoding method in our quantum memory compiler.
Figure~\ref{QuantumCost} illustrates the quantum cost for each encoding (including in optimized form) when using what we term a `natural' gate set for each encoding.
The `natural' gate set for basis encoding consists of Pauli-$\mathbf{X}$ gates and two-controlled (\textit{i.e.,} ``true''--not generalized) Toffoli gates; for angle and improved angle encoding, it consists of multi-controlled rotation gates and single rotation gates.
Figure~\ref{DecomposedQuantumCost} illustrates the same cost, but for a uniform gate set for all encoding approaches.
We chose a uniform gate set that corresponds to an implementable gate set on IBM's suite of quantum devices: $\mathbf{R}_x$, $\mathbf{R}_y$, $\mathbf{R}_z$, and controlled-Pauli-$\mathbf{X}$ gates.
Figure~\ref{QuantumDepth} illustrates the quantum depth of circuits for each encoding method and optimization, and Figure~\ref{Qubits} does the same for the number of qubits.
Before discussing trends in these results, we note that tables containing the numerical data from which the figures were generated can be found in the Appendix (Sec.~\ref{Appendix}).
Additionally, each figure contains a graphical legend that is described in each caption.

First, we consider the effect of optimization.
For all encodings, optimization has an important role, even as the available optimizations differ by encoding type.
As illustrated in Figures~\ref{QuantumCost} and ~\ref{DecomposedQuantumCost}, the optimized versions of the encoded circuits offer markedly lower costs, both when the `natural' and decomposed gate sets are used.
On average, `natural' gate set quantum cost decreases by 51\% when appropriate optimization is applied to each encoded circuit.
We see similar results for quantum depth in Figure~\ref{QuantumDepth}, which illustrates that---again, on average---applying an appropriate optimization decreases depth by 66\%.

Although optimization improves the cost and depth for all circuits considered, we note that the amount of benefit depends upon encoding type: the optimization used for basis-encoded circuits (double-NOT gate removal) has far less effect upon circuit performance than does the gray-code optimization applied to circuits encoded using angle and improved angle encoding.
Indeed, while the average `natural' gate set quantum cost reduction for basis-encoded circuits is merely 6\%, for angle- and improved-angle-encoded circuits, it is 81\% and 66\%, respectively.
The effect is even more dramatic for quantum depth: while double-NOT removal improves depth for basis-encoded circuits by an average of 0\%, gray-code optimization reduces the average quantum depth of angle- and improved angle-encoded circuits by 99\% in both cases.

Before turning to a comparison of the encodings themselves, we make three further observations.
First, without the gray-code optimization, angle and improved angle encoding are often as expensive---or even more so---in terms of cost and depth than basis encoding; however, when optimization is applied, cost and depth drop so dramatically that they are often better.
This illustrates one reason that further study of different encodings---including further development of optimizations---is of such importance.
Second and relatedly, optimization has a dramatic role upon the number of gates required to represent the encoded circuits using a uniform gate set, which we chose to be a gate set suitable for some existing quantum hardware.
In particular, we note that---for all of the memory images we considered---decomposing optimized versions of the angle- and improved-angle-encoded circuits requires a cost increase of one order of magnitude, while decomposing non-optimized versions results in a cost increase of two or more orders of magnitude.
In today's era of NISQ hardware, this illustrates the importance of choosing encodings that can be optimized as much as possible.
Third and finally, we note that none of the applied optimizations are designed to decrease qubit count, which is why there is no effect of optimization on qubit count.

After considering the effects of optimization, we can directly compare the encoding types.
We will do so using the optimized versions of the encodings, since those represent the best that each encoding has to offer in this work. 
We first consider quantum cost: basis encoding is often more expensive than angle encoding.
On average, angle-encoded circuits have a cost 63\% lower than that for basis-encoded circuits for the same memory images.
The improvement for improved angle encoding is naturally less because improved angle encoding involves storing two sets of values (mantissas and exponents), which require more gates.
(Recall that the improvements of improved angle encoding are precision of storage and fault tolerance, not number of operations.)
And yet, it is thus quite notable that there is still cost improvement when compared to basis encoding: on average, improved-angle-encoded circuits have a cost 29\% lower than that for basis-encoded circuits for the same memory images.

The same trend holds true for quantum depth: switching to angle or improved angle encoding from basis encoding yields an average depth decrease of 44\% or 38\%, respectively.
This is a particularly important finding for NISQ machines, because quantum depth is an important way of assessing whether circuits will likely be able to run to completion without noise entirely deteriorating the results~\cite{preskill2018quantum}.
Thus, given the ability of angle and improved angle encoding to dramatically improve quantum depth, future work ought to consider how algorithms currently relying upon basis encoding might adapt to take advantage of a different data encoding that could be more usable on near-term hardware.
(We consider one example at the end of this section.)

Finally, we consider the effect of encoding upon required qubits.
Again, basis encoding is generally quite a bit more expensive in terms of qubits.
While basis encoding requires at least one qubit for each of the memory image's input and output bits, it can require even more---as it does in all circuits synthesized here---given the breakdown of generalized Toffoli gates into two-controlled Toffoli gates.
It is important to consider this breakdown, because current quantum hardware requires such decomposition to run circuits involving Toffoli gates~\cite{monz2009realization,fedorov2012implementation}.
So, it is noteworthy that, on average, angle- and improved angle-encoded circuits have 61\% fewer qubits than basis-encoded circuits.
Again, for NISQ-era hardware, where qubits are relatively few and where even existing qubits are often prone to error, reducing the number required to perform any task---including to store data---is of critical import.
It is also worth noting that this benefit becomes more stark the larger the size of the data that needs storing: because basis encoding requires one qubit for each bit in that data, the number of required qubits grows linearly with the size of the data words to be stored.
Conversely, both angle and improved angle encoding require just a single qubit to store the data, regardless of the size of the encoded data words.

All of these results suggest that algorithms using databases encoded with basis encoding may be needlessly costly in terms of quantum cost, quantum depth, and qubit count.
For example, consider Grover's algorithm, which searches a database for a specified value or values.
Grover's algorithm involves an ``oracle'' that is a database-dependent portion of the circuit that uses basis encoding to represent the database to be searched.
Given the possibility for lower quantum cost, depth, and qubit count, we have begun to explore the algorithmic ramifications when we attempt to synthesize the oracle using angle encoding, improved angle encoding, and other encoding approaches not described in this work.
The portions of the algorithm acting upon the data would, of course, have to be modified to obtain it from a differently-encoded database, and determining the effects of those changes, including whether they result in an overall lower quantum cost, depth, or qubit count, both for Grover's algorithm and other quantum algorithms that depend upon oracles functions is an ongoing research effort.  Since oracle functions can be specified in a tabular form, such as a \texttt{.pla} file, and can be synthesized using the same methods as those for QROM and with various encoding approaches, all of the results described here apply equally to quantum circuits that utilize oracle or other custom functions.

\section{Conclusion and Future Work}
In this paper, we have proposed a standard definition for QROM and three data encoding methods, described a new encoding method and optimization thereof, and provided experimental results for an automated memory compilation tool that generates QROM state generation circuits with any of the three described encoding approaches.
We interpret a set of benchmark functions described as \texttt{.pla} files as memory images that show the effect of encoding type on quantum circuit metrics.
We find that, as expected, the encoding type significantly affects the characteristics of the QROM circuit, making the decision of which encoding to use an important one.
In particular, future work will consider the effect of encoding type in specific applications, and not solely upon benchmark circuits, as were considered here.
Our automated tool for generating and optimizing QROMs with various encoding approaches provides an efficient and effective way to generate encoded circuits that eases the tasks required of a QC algorithm designer.

\bibliographystyle{IEEEtran}
\bibliography{biblio}

\section{Appendix}\label{Appendix}
The following tables present circuit metrics for twelve benchmark circuits, each of which was synthesized using six procedures: basis encoding, angle encoding, improved angle encoding, and an optimized version of each.
Specifically, the optimization applied for basis-encoded circuits is double-NOT gate removal, and, for angle and improved angle-encoded circuits, is the gray-code optimization described in Sec.~\ref{Gray_Code_Optimization}.
For each circuit, two sets of metrics are presented: the gate count and quantum cost for the most ``natural'' gate set for each encoding, and the gate count, quantum cost, and quantum depth for a uniform gate set native to the IBMQ suite of quantum devices.
(Qubit count is also presented, and does not differ by gate set.)
In particular, the ``natural'' gate set for basis encoding is Pauli-$\mathbf{X}$ gates and two-controlled (\textit{i.e.,} ``true'') Toffoli gates, and the same for the angle and improved angle encoding approaches is multi-controlled rotation gates and single rotation gates.
The uniform gate set is the set of single rotation gates and the Controlled-NOT gate; to decompose circuits into this format---and to collect results from the decomposed circuit---we integrated calls to some Qiskit functionality from MustangQ, which allowed for automating the decomposition.

\begin{table*}[htbp]
\caption{Data for benchmark circuit \texttt{squar5}.\label{squar5}}
\begin{center}
\begin{tabular}{|c|c|c|c|c|c|c|c|c|c|}\hline
Encoding & Optimized & Inputs & Outputs & Qubits & Gate Count & Quantum Cost & \multicolumn{1}{|p{2cm}|}{\centering Decomposed \\ Gate Count} & \multicolumn{1}{|p{2cm}|}{\centering Decomposed \\ Quantum Cost} & \multicolumn{1}{|p{2cm}|}{\centering Decomposed \\ Quantum Depth} \\
\hline
Basis & No & 5 & 8 & 17 & 152 & 373 & 1800 & 2433 & 1092 \\
Basis & Yes & 5 & 8 & 17 & 134 & 355 & 1782 & 2415 & 1092 \\
Angle & No & 5 & 8 & 6 & 172 & 322 & 14992 & 17752 & 12092 \\ 
Angle & Yes & 5 & 8 & 6 & 64 & 96 & 512 & 576 & 448 \\
Improved Angle & No & 5 & 8 & 6 & 207 & 402 & 19473 & 23061 & 15719 \\
Improved Angle & Yes & 5 & 8 & 6 & 128 & 192 & 576 & 672 & 512 \\
\hline
\end{tabular}
\end{center}
\end{table*}

\begin{table*}[htbp]
\caption{Data for benchmark circuit \texttt{Z9sym}.\label{Z9sym}}
\begin{center}
\begin{tabular}{|c|c|c|c|c|c|c|c|c|c|}\hline
Encoding & Optimized & Inputs & Outputs & Qubits & Gate Count & Quantum Cost & \multicolumn{1}{|p{2cm}|}{\centering Decomposed \\ Gate Count} & \multicolumn{1}{|p{2cm}|}{\centering Decomposed \\ Quantum Cost} & \multicolumn{1}{|p{2cm}|}{\centering Decomposed \\ Quantum Depth} \\
\hline
Basis & No & 9 & 1 & 18 & 1065 & 2407 & 11401 & 15327 & 7257 \\
Basis & Yes & 9 & 1 & 18 & 803 & 2145 & 11139 & 15065 & 7257 \\
Angle & No & 9 & 1 & 10 & 4200 & 7980 & 3437280 & 4080720 & 2790061 \\ 
Angle & Yes & 9 & 1 & 10 & 768 & 1280 & 7936 & 8960 & 6912 \\
Improved Angle & No & 9 & 1 & 10 & 8400 & 15960 & 6874560 & 8161440 & 5580121 \\
Improved Angle & Yes & 9 & 1 & 10 & 1536 & 2560 & 8704 & 10240 & 7680 \\
\hline
\end{tabular}
\end{center}
\end{table*}

\begin{table*}[htbp]
\caption{Data for benchmark circuit \texttt{inc}.\label{inc}}
\begin{center}
\begin{tabular}{|c|c|c|c|c|c|c|c|c|c|}\hline
Encoding & Optimized & Inputs & Outputs & Qubits & Gate Count & Quantum Cost & \multicolumn{1}{|p{2cm}|}{\centering Decomposed \\ Gate Count} & \multicolumn{1}{|p{2cm}|}{\centering Decomposed \\ Quantum Cost} & \multicolumn{1}{|p{2cm}|}{\centering Decomposed \\ Quantum Depth} \\
\hline
Basis & No & 7 & 9 & 22 & 702 & 1787 & 8814 & 11927 & 5782 \\
Basis & Yes & 7 & 9 & 22 & 600 & 1685 & 8712 & 11825 & 5774 \\
Angle & No & 7 & 9 & 8 & 896 & 1624 & 212016 & 251536 & 171706 \\ 
Angle & Yes & 7 & 9 & 8 & 256 & 384 & 2048 & 2304 & 1792 \\
Improved Angle & No & 7 & 9 & 8 & 972 & 1756 & 228332 & 270892 & 184914 \\
Improved Angle & Yes & 7 & 9 & 8 & 702 & 1787 & 8814 & 11927 & 1960 \\
\hline
\end{tabular}
\end{center}
\end{table*}

\begin{table*}[htbp]
\caption{Data for benchmark circuit \texttt{Z5xp1}.\label{Z5xp1}}
\begin{center}
\begin{tabular}{|c|c|c|c|c|c|c|c|c|c|}\hline
Encoding & Optimized & Inputs & Outputs & Qubits & Gate Count & Quantum Cost & \multicolumn{1}{|p{2cm}|}{\centering Decomposed \\ Gate Count} & \multicolumn{1}{|p{2cm}|}{\centering Decomposed \\ Quantum Cost} & \multicolumn{1}{|p{2cm}|}{\centering Decomposed \\ Quantum Depth} \\
\hline
Basis & No & 7 & 10 & 23 & 412 & 981 & 4524 & 6121 & 2897 \\
Basis & Yes & 7 & 10 & 23 & 351 & 921 & 4464 & 6061 & 2895 \\
Angle & No & 7 & 10 & 8 & 1024 & 1920 & 260864 & 309504 & 211330 \\ 
Angle & Yes & 7 & 10 & 8 & 256 & 384 & 2048 & 2304 & 1792 \\
Improved Angle & No & 7 & 10 & 8 & 1157 & 2228 & 311747 & 369887 & 252605 \\
Improved Angle & Yes & 7 & 10 & 8 & 512 & 768 & 2304 & 2688 & 2048 \\
\hline
\end{tabular}
\end{center}
\end{table*}

\begin{table*}[htbp]
\caption{Data for benchmark circuit \texttt{dist}.\label{dist}}
\begin{center}
\begin{tabular}{|c|c|c|c|c|c|c|c|c|c|}\hline
Encoding & Optimized & Inputs & Outputs & Qubits & Gate Count & Quantum Cost & \multicolumn{1}{|p{2cm}|}{\centering Decomposed \\ Gate Count} & \multicolumn{1}{|p{2cm}|}{\centering Decomposed \\ Quantum Cost} & \multicolumn{1}{|p{2cm}|}{\centering Decomposed \\ Quantum Depth} \\
\hline
Basis & No & 8 & 5 & 20 & 1562 & 3952 & 19770 & 26712 & 12789 \\
Basis & Yes & 8 & 5 & 20 & 1348 & 3738 & 19556 & 26498 & 12781 \\
Angle & No & 8 & 5 & 9 & 2287 & 4327 & 1042177 & 1236997 & 845326 \\ 
Angle & Yes & 8 & 5 & 9 & 512 & 768 & 4096 & 4608 & 3584 \\
Improved Angle & No & 8 & 5 & 9 & 2588 & 5052 & 1258612 & 1493924 & 1021021 \\
Improved Angle & Yes & 8 & 5 & 9 & 1024 & 1536 & 4608 & 5376 & 4096 \\
\hline
\end{tabular}
\end{center}
\end{table*}

\begin{table*}[htbp]
\caption{Data for benchmark circuit \texttt{f51m}.\label{f51m}}
\begin{center}
\begin{tabular}{|c|c|c|c|c|c|c|c|c|c|}\hline
Encoding & Optimized & Inputs & Outputs & Qubits & Gate Count & Quantum Cost & \multicolumn{1}{|p{2cm}|}{\centering Decomposed \\ Gate Count} & \multicolumn{1}{|p{2cm}|}{\centering Decomposed \\ Quantum Cost} & \multicolumn{1}{|p{2cm}|}{\centering Decomposed \\ Quantum Depth} \\
\hline
Basis & No & 8 & 8 & 22 & 352 & 797 & 3504 & 4737 & 2249 \\
Basis & Yes & 8 & 8 & 22 & 284 & 729 & 3436 & 4669 & 2245 \\
Angle & No & 8 & 8 & 9 & 2295 & 4335 & 1042185 & 1237005 & 845327 \\ 
Angle & Yes & 8 & 8 & 9 & 512 & 768 & 4096 & 4608 & 3584 \\
Improved Angle & No & 8 & 8 & 9 & 3421 & 6485 & 1565295 & 1857907 & 1269647 \\
Improved Angle & Yes & 8 & 8 & 9 & 896 & 1408 & 4480 & 5248 & 3968 \\
\hline
\end{tabular}
\end{center}
\end{table*}

\begin{table*}[htbp]
\caption{Data for benchmark circuit \texttt{mlp4}.\label{mlp4}}
\begin{center}
\begin{tabular}{|c|c|c|c|c|c|c|c|c|c|}\hline
Encoding & Optimized & Inputs & Outputs & Qubits & Gate Count & Quantum Cost & \multicolumn{1}{|p{2cm}|}{\centering Decomposed \\ Gate Count} & \multicolumn{1}{|p{2cm}|}{\centering Decomposed \\ Quantum Cost} & \multicolumn{1}{|p{2cm}|}{\centering Decomposed \\ Quantum Depth} \\
\hline
Basis & No & 8 & 8 & 23 & 949 & 2494 & 12693 & 5196 & 8206 \\
Basis & Yes & 8 & 8 & 23 & 865 & 2410 & 12609 & 17090 & 8206 \\
Angle & No & 8 & 8 & 9 & 1905 & 3705 & 919455 & 1091355 & 745876 \\ 
Angle & Yes & 8 & 8 & 9 & 512 & 768 & 1050119 & 1246467 & 851956 \\
Improved Angle & No & 8 & 8 & 9 & 2073 & 4129 & 1050119 & 1246467 & 851956 \\
Improved Angle & Yes & 8 & 8 & 9 & 844 & 1356 & 4428 & 5196 & 3916 \\
\hline
\end{tabular}
\end{center}
\end{table*}

\begin{table*}[htbp]
\caption{Data for benchmark circuit \texttt{clip}.\label{clip}}
\begin{center}
\begin{tabular}{|c|c|c|c|c|c|c|c|c|c|}\hline
Encoding & Optimized & Inputs & Outputs & Qubits & Gate Count & Quantum Cost & \multicolumn{1}{|p{2cm}|}{\centering Decomposed \\ Gate Count} & \multicolumn{1}{|p{2cm}|}{\centering Decomposed \\ Quantum Cost} & \multicolumn{1}{|p{2cm}|}{\centering Decomposed \\ Quantum Depth} \\
\hline
Basis & No & 9 & 5 & 22 & 1379 & 3242 & 15635 & 21062 & 10011 \\
Basis & Yes & 9 & 5 & 22 & 1099 & 2962 & 15355 & 20782 & 1011 \\
Angle & No & 9 & 5 & 10 & 4956 & 9420 & 4059260 & 4819132 & 3294929 \\ 
Angle & Yes & 9 & 5 & 10 & 1024 & 1536 & 8192 & 9216 & 7168 \\
Improved Angle & No & 9 & 5 & 10 & 9093 & 17382 & 7537347 & 8948319 & 6118204 \\
Improved Angle & Yes & 9 & 5 & 10 & 2048 & 3072 & 9216 & 10752 & 8192 \\
\hline
\end{tabular}
\end{center}
\end{table*}

\begin{table*}[htbp]
\caption{Data for benchmark circuit \texttt{addm4}.\label{add_m4}}
\begin{center}
\begin{tabular}{|c|c|c|c|c|c|c|c|c|c|}\hline
Encoding & Optimized & Inputs & Outputs & Qubits & Gate Count & Quantum Cost & \multicolumn{1}{|p{2cm}|}{\centering Decomposed \\ Gate Count} & \multicolumn{1}{|p{2cm}|}{\centering Decomposed \\ Quantum Cost} & \multicolumn{1}{|p{2cm}|}{\centering Decomposed \\ Quantum Depth} \\
\hline
Basis & No & 9 & 8 & 25 & 3683 & 8922 & 43875 & 59162 & 28278 \\
Basis & Yes & 9 & 8 & 25 & 3005 & 8244 & 43197 & 58484 & 28278 \\
Angle & No & 9 & 8 & 10 & 4702 & 9022 & 3928222 & 4663582 & 3188641 \\
Angle & Yes & 9 & 8 & 10 & 1024 & 1536 & 8192 & 9216 & 7168 \\
Improved Angle & No & 9 & 8 & 10 & 4731 & 9078 & 3952773 & 4692729 & 3208570 \\
Improved Angle & Yes & 9 & 8 & 10 & 2042 & 3066 & 9210 & 10746 & 8186 \\
\hline
\end{tabular}
\end{center}
\end{table*}

\begin{table*}[htbp]
\caption{Data for benchmark circuit \texttt{b11}.\label{b11}}
\begin{center}
\begin{tabular}{|c|c|c|c|c|c|c|c|c|c|}\hline
Encoding & Optimized & Inputs & Outputs & Qubits & Gate Count & Quantum Cost & \multicolumn{1}{|p{2cm}|}{\centering Decomposed \\ Gate Count} & \multicolumn{1}{|p{2cm}|}{\centering Decomposed \\ Quantum Cost} & \multicolumn{1}{|p{2cm}|}{\centering Decomposed \\ Quantum Depth} \\
\hline
Basis & No & 8 & 31 & 44 & 416 & 1031 & 4848 & 6571 & 3126 \\
Basis & Yes & 8 & 31 & 44 & 364 & 979 & 4796 & 6519 & 3122 \\
Angle & No & 8 & 31 & 9 & 2304 & 4352 & 1046272 & 1241856 & 848642 \\ 
Angle & Yes & 8 & 31 & 9 & 320 & 576 & 3904 & 4416 & 3392 \\
Improved Angle & No & 8 & 31 & 9 & 2304 & 4352 & 1046272 & 1241856 & 848642 \\
Improved Angle & Yes & 8 & 31 & 9 & 576 & 1088 & 4160 & 4928 & 3648 \\
\hline
\end{tabular}
\end{center}
\end{table*}

\begin{table*}[htbp]
\caption{Data for benchmark circuit \texttt{apex4}.\label{apex4}}
\begin{center}
\begin{tabular}{|c|c|c|c|c|c|c|c|c|c|}\hline
Encoding & Optimized & Inputs & Outputs & Qubits & Gate Count & Quantum Cost & \multicolumn{1}{|p{2cm}|}{\centering Decomposed \\ Gate Count} & \multicolumn{1}{|p{2cm}|}{\centering Decomposed \\ Quantum Cost} & \multicolumn{1}{|p{2cm}|}{\centering Decomposed \\ Quantum Depth} \\
\hline
Basis & No & 9 & 19 & 36 & 61556 & 172859 & 907908 & 1230799 & 598257 \\
Basis & Yes & 9 & 19 & 36 & 58784 & 170087 & 905136 & 1228027 & 598255 \\
Angle & No & 9 & 19 & 10 & 5120 & 9728 & 4190208 & 4974592 & 3401218 \\
Angle & Yes & 9 & 19 & 10 & 1024 & 1536 & 8192 & 9216 & 7168 \\
Improved Angle & No & 9 & 19 & 10 & 5120 & 9728 & 4190208 & 4974592 & 3401218 \\
Improved Angle & Yes & 9 & 19 & 10 & 2560 & 1536 & 8704 & 10240 & 7680 \\
\hline
\end{tabular}
\end{center}
\end{table*}

\begin{table*}[htbp]
\caption{Data for benchmark circuit \texttt{ex5}.\label{ex5}}
\begin{center}
\begin{tabular}{|c|c|c|c|c|c|c|c|c|c|}\hline
Encoding & Optimized & Inputs & Outputs & Qubits & Gate Count & Quantum Cost & \multicolumn{1}{|p{2cm}|}{\centering Decomposed \\ Gate Count} & \multicolumn{1}{|p{2cm}|}{\centering Decomposed \\ Quantum Cost} & \multicolumn{1}{|p{2cm}|}{\centering Decomposed \\ Quantum Depth} \\
\hline
Basis & No & 8 & 63 & 78 & 7024 & 19525 & 101920 & 138145 & 67128 \\
Basis & Yes & 8 & 63 & 78 & 6678 & 19179 & 101574 & 137799 & 67128 \\
Angle & No & 8 & 63 & 9 & 2304 & 4352 & 1046272 & 1241856 & 848642 \\ 
Angle & Yes & 8 & 63 & 9 & 512 & 768 & 4096 & 4608 & 3584 \\
Improved Angle & No & 8 & 63 & 9 & 2688 & 4992 & 1177152 & 1397184 & 954722 \\
Improved Angle & Yes & 8 & 63 & 9 & 776 & 1288 & 4360 & 5128 & 3848 \\
\hline
\end{tabular}
\end{center}
\end{table*}

\end{document}